\documentclass[reprint, amsmath,amssymb,aps,pra,longbibliography, superscriptaddress]{revtex4-2}

\usepackage{graphicx}
\usepackage{dcolumn}
\usepackage{bm}
\usepackage{hyperref}
\usepackage{siunitx}
\usepackage[table]{xcolor}
\usepackage[caption=false]{subfig}
\newcolumntype{P}[1]{>{\centering\arraybackslash}p{#1}} 
\usepackage{mathtools}

\DeclarePairedDelimiter\ket{\lvert}{\rangle}
\DeclarePairedDelimiterX\braket[2]{\langle}{\rangle}{#1\,\delimsize\vert\,\mathopen{}#2}

\usepackage{letltxmacro}
\LetLtxMacro{\originaleqref}{\eqref}
\renewcommand{\eqref}{Eq.~\originaleqref}

\definecolor{dark-gray}{gray}{0.40}
\definecolor{quantumviolet}{HTML}{53257F} 
\definecolor{quantumlightviolet}{HTML}{A088B1}
\definecolor{quantumgreen}{HTML}{00826F}
\definecolor{quantumrose}{HTML}{EDB3FF} 
\definecolor{quantumdarkrose}{HTML}{F06292}
\definecolor{quantumturquoise}{HTML}{00C9AF}
\definecolor{quantumblue}{HTML}{85B1CC}
\definecolor{quantumdarkgray}{HTML}{4C4452}
\definecolor{quantumgray}{HTML}{555555}
\definecolor{black}{HTML}{000000}

\hypersetup{colorlinks=true, 
            linkcolor=black,
            citecolor=black,
            urlcolor=black
}

\begin{document}

\title{Demonstration of quantum projective simulation \texorpdfstring{\\ on a single-photon-based quantum computer}{}}

\author{Giacomo Franceschetto}
\email{giacomo.franceschetto@icfo.eu}
\affiliation{Quandela, 7 Rue Léonard de Vinci, 91300 Massy, France}
\affiliation{ICFO-Institut de Ciencies Fotoniques, The Barcelona Institute of Science and Technology, Av. Carl Friedrich Gauss 3, 08860 Castelldefels (Barcelona), Spain}

\author{Arno Ricou}
\affiliation{Quandela, 7 Rue Léonard de Vinci, 91300 Massy, France}

\begin{abstract}
Variational quantum algorithms show potential in effectively operating on noisy intermediate-scale quantum devices. A novel variational approach to reinforcement learning has been recently proposed, incorporating linear-optical interferometers and a classical learning model known as projective simulation (PS). PS is a decision-making tool for reinforcement learning and can be classically represented as a random walk on a graph that describes the agent’s memory. In its optical quantum version, this approach utilizes quantum walks of single photons on a mesh of tunable beamsplitters and phase shifters to select actions. In this work, we present the implementation of this algorithm on Ascella, a single-photon-based quantum computer from Quandela. The focus is drawn on solving a test bed task to showcase the potential of the quantum agent with respect to the classical agent.
\end{abstract}
\date{\today}

\maketitle

\section{Introduction}
Variational quantum algorithms (VQAs) have emerged as a promising class of algorithms for leveraging the computational capabilities of noisy intermediate-scale quantum devices. These algorithms utilize parameterized quantum circuits to perform tasks by optimizing parameters through classical feedback loops, making them adaptable to the error-prone nature of current quantum hardware \cite{cerezo_variational_2021}. The flexibility and hybrid quantum-classical nature of VQAs have made them a focal point in the exploration of quantum computing applications, particularly in the domain of quantum machine learning (QML) \cite{benedetti_parameterized_2019, cerezo_challenges_2022, biamonte_quantum_2017}. 

One of the objectives of QML is to harness quantum effects to enhance performance in specific tasks compared to traditional methods. However, this must be very carefully done since simply incorporating quantum elements into a learning model does not necessarily make it hard to approximate using classical methods \cite{landman_classically_2022} or better than a classical model \cite{bowles_better_2024}. In addition, the introduction of quantum effects into machine learning algorithms often complicates even more the interpretation of their results. To address this last issue, a variational method \cite{Flamini_2024} was recently introduced to quantize the projective simulation model \cite{briegel_projective_2012}, which is a reinforcement learning algorithm valued for its interpretability. While the classical version of PS has been applied and benchmarked in various scenarios \cite{mautner_projective_2015, melnikov_projective_2014}, its quantum counterpart has remained within the realm of theoretical studies \cite{paparo_quantum_2014}, with only some proposals of experimental realizations. Namely, experimental protocols to realize the fundamental working principles of quantum PS have been proposed in photonic \cite{flamini_photonic_2020, Flamini_2024}, ion-trap \cite{dunjko_quantum-enhanced_2015} and superconducting platforms \cite{friis_coherent_2015, lamata_basic_2017}. Some of these working principles have been demonstrated experimentally \cite{sriarunothai2018speeding}. However, to the best of the author's knowledge, there has not yet been a full implementation of a quantum PS agent on quantum hardware solving a task.

This work aims to bridge this gap by implementing a variational quantum PS algorithm on a photonic platform. The platform selection is especially advantageous since, in light of the most recent theoretical proposal \cite{Flamini_2024}, no additional technological advancement is required, as the current state-of-the-art single-photon-based quantum computers already fulfill all the hardware prerequisites. Additionally, we evaluate our implementation of the quantum PS agent using a modified version of a learning scenario that was previously introduced \cite{Flamini_2024, ried_how_2019}. This test bed task is designed to create an environment in which correct decision-making by the PS agent is conditional on the use of quantum effects. Our testing includes ideal and noisy simulations, as well as execution on Quandela's single-photon-based quantum computer \cite{maring_versatile_2024}.
\begin{figure}[b!]
    \includegraphics{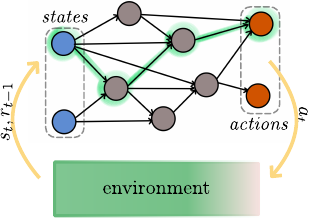}
    \caption{\label{fig:ps} \textbf{Projective simulation for reinforcement learning.} The agent is modeled as a weighted directed graph. Blue vertices represent states, while orange vertices represent possible actions. The highlighted path in green is an example of a deliberation process.}
\end{figure}
\section{Background}
\subsection{Classical projective simulation }
In reinforcement learning (RL), an agent interacts with the environment to learn an optimal decision-making strategy.
During each interaction, the environment provides the agent with its current state and a reward associated with the previous interaction. The agent must select an action from a set of options and send it back to the environment. The optimal strategy is the one that maximizes the reward accumulated over the long term. PS employs a random walk over a weighted directed graph called episodic compositional memory (ECM) to make decisions. Each vertex of the graph, referred to as a clip, represents a memory unit of the agent. The first layer of clips corresponds to states, while the last layer corresponds to actions, following the order of the graph. The deliberation is governed by a stochastic process that starts with the input state clip and ends with one of the action clips (Fig. \ref{fig:ps}). 
After each interaction, the weights of the edges involved in the walk are updated based on the obtained reward:

\begin{equation}\label{classical_update_rule}
    h^{t+1} = h^{t} + r^t,
\end{equation}
where $h^t$ is the collection of weights of the edges involved in the walk at time $t$ and $r^t$ is the reward at time $t$. The reward is the same for all the weights and can be either positive and then increase the weights, or negative and then decrease them. The weight of the edge between two clips $i, j$ is connected to the jump probability from clip $i$ to clip $j$:
\begin{equation}
    p_{ij} = \frac{h_{ij}}{\sum_{j'}h_{ij'}}.
\end{equation}
More advanced update rules can be introduced, e.g. to account for forgetting and glowing mechanisms \cite{mautner_projective_2015}. 

The main distinction between PS and other decision-making algorithms for RL is that PS employs the ECM to model the agent. This not only enables higher level of abstraction within deliberations compared to tabular approaches such as Q-Learning and SARSA \cite{sutton_reinforcement_2018}, but also enriches the agent by making its decisions interpretable, which is a significant concern in deep reinforcement learning models \cite{annasamy_towards_2019}.

\subsection{Quantum optical projective simulation }
\begin{figure}[ht]
\includegraphics{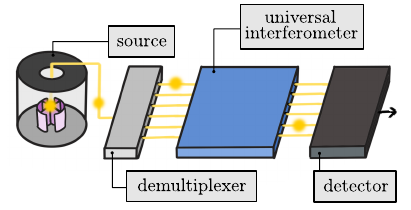}
\caption{\label{fig:opticalsetup} \textbf{Quantum optical setup.} This scheme is a basic description of the hardware considered in this work which is the single-photon-based quantum computer Ascella \cite{maring_versatile_2024}.}
\end{figure}

The natural quantization of PS involves replacing the random walk through the ECM with a quantum walk over a quantum model of the agent's memory \cite{briegel_projective_2012}. The quantum model, in its optical version, consists of a single photon passing through a quantum optical setup (Fig. \ref{fig:opticalsetup}). 
Each optical mode, denoted as $a_i$, corresponds to a vertex of the classical ECM. 
Let $a_k$ and $a_l$ be optical modes corresponding to a state clip and an action clip, respectively. 
The decision-making process for a photon injected at input port $k$ and detected at output port $l$ is modeled by:
\begin{equation} \label{eq:qps}
    \hat{a_l}^\dagger_{\text{out}} = U_{lk}^{\text{ECM}}\hat{a_k}^\dagger_{\text{in}},
\end{equation}
where $U^{\text{ECM}}$ is the realization of such process by the interferometric scattering matrix connecting the $M$ different optical modes.

While the classical ECM provides access to intermediate clip-to-clip transition probabilities as they constitute the graph's weights, this accessibility is not possible within the quantum PS framework. Given a photon in the \emph{state} mode $a_k$ as input, only the probability of getting the photon in the output \emph{action} mode $a_l$ can be measured. In this case, a variational approach can be used to update the quantum ECM after each interaction, with the goal of finding the minimum of a loss function that embeds an update rule similar to \eqref{classical_update_rule}, but for input/output modes probabilities instead of intermediate weights:
\begin{equation}\label{loss_fn}
    \mathcal{L}_{PS}(\vec{\theta}) = D[p_{kl}^{t+1}(\vec{\theta}), C(p_{kl}^{t}+r^t)],
\end{equation}
where $\vec{\theta}$ is the vector of parameters of the quantum ECM, $D$ is a suitable distance function and $C$ is a cutoff function used to clamp its argument to a value in the $[0, 1]$ range. In \eqref{loss_fn}, the probability at time step $t$ is not dependent on the circuit parameters, as it is considered a fixed value that was previously computed. Instead, at time step $t+1$ the objective is to tune the parameters in order to change the probability, thereby introducing a dependence.

One motivation to formulate a quantum version of PS is based on the possibility of enhancing the agent's decision-making capabilities thanks to quantum effects. In classical PS, the stochastic process through the clips of the ECM that leads the agent from the received state to an action has a well-defined trajectory. In the quantum version, the trajectory of the single photon within the optical setup can be delocalized over multiple optical modes. The quantum agent's deliberation process involves the photon exploring multiple modes in a superposition state as it passes through the optical interferometer. Moreover, the quantum agent can learn to exploit superposition states using interference. This quantum approach introduces a set of deliberation patterns that the classical counterpart may not achieve, thereby enabling the agent to solve specific tasks differently, and potentially more efficiently. One aspect of the quantized model that is not straightforward to understand is the inheritance of the interpretability of the classical PS. This aspect is one of the main focuses explored in \cite{Flamini_2024}, where the notion of partial traceability is introduced with the goal of partially recovering some interpretability also in the quantum framework.

\section{Learning scenario } 
The learning scenario presented here was first introduced by  \cite{ried_how_2019}. Then, it was reformulated in a transfer learning setting to provide an example task that a quantum PS agent can solve efficiently, but its classical counterpart cannot \cite{Flamini_2024}.  The same transfer learning setting is used here with some simplifications to enable hardware implementation.  

The task in question is to predict the result of an experiment $\mathcal{E}$ on a finite set of percepts $\{\mathcal{P}_i\}$. The percept in this framework can be intended as the state provided by the environment. 
Here, each percept $\mathcal{P}_i$ is fully characterized by two observables (\emph{color} and \emph{shape}), and each observable has two possible values (\emph{red, blue} and \emph{circle, square}). This results in a total of four percepts covering all possible combinations of observable values. We assume that each percept is equally likely to be tested. The experiment is a “yes"/“no" question about the two observables. Of the four different possibilities, we pick \emph{is the percept both blue and a square?}; in other words, we want the agent to learn to detect the target percept \emph{blue-square}. 
The simplifications made to the task described in \cite{Flamini_2024} consist in reducing both the number of observables and the possible values that each observable can take from three to two, and in considering only one experiment rather than all possible ones.

\begin{figure*}[t!]
\includegraphics[width=0.8\textwidth]{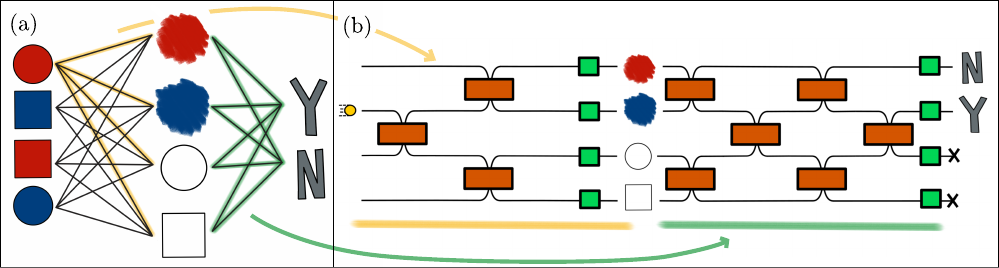}
\caption{\label{fig:qps} \textbf{Classical and quantum implementation of the ECM for the transfer learning scenario.} (a) The classical version of the full ECM is a three-layer graph, with only the first two layers considered in stage 1 and the full graph considered in stage 2. (b) In the quantum version, the ECM is created using a network of tunable beam splitters (rectangle elements) and phase shifters (square elements). The circuit underlined in yellow in stage 1 realizes the quantum ECM for a single percept, with different sets of parameters considered for different percepts. In stage 2, we take both underlined circuits in sequence. We make the encoding of optical modes to actions explicit with the illustration.}
\end{figure*}

In the transfer learning setting, the task consists of two stages. In stage 1, the agent learns the correct observable values for each percept. This corresponds to training the ECM on the first two layers of the graph shown in Fig. \ref{fig:qps} (a). In stage 2, the agent uses the knowledge of the observables from the previous stage to predict the experiment's answer. This corresponds to training the ECM on the full graph shown in Fig. \ref{fig:qps} (a).

To solve the task with $100\%$ accuracy, ones need simultaneous knowledge of the two observables. The trained classical PS agent can only use the knowledge of one of the observables at a time, i.e. it can only take one localized path at a time when choosing an action. Consequently, the optimal classical strategy for the agent is to always respond negatively for the percept that has no observable values in common with the target one, to respond negatively $75\%$ of the times for the two percepts with one overlapping observable value, and to respond positively or negatively at random for the target percept. This leads to a maximum accuracy of $75\%$ for the classical agent \cite{A_A}. The quantum agent, on the other hand, is not limited in its maximum accuracy. Thanks to the non-local decision path, it is possible to access the knowledge of both observables within the same deliberation, and full accuracy can be achieved.

\section{Architecture } 
The standard approach proposed in Ref. \cite{Flamini_2024} to implement quantum PS agents consists in reproducing the scattering matrix $U^{ECM}$ of \eqref{eq:qps} as a set of unitaries $\{U^k\}$, one for each vertex of the classical ECM graph. Each of the unitaries $U^k$ can be realized by tuning a mesh of beam splitters and phase shifters arranged in a universal scheme \cite{reck_experimental_1994, clements_optimal_2016}. 

In stage 1 of the learning scenario we are considering, the goal is to map the input percept states to the correct observable-value states. Given the current encoding of the task information into the optical setup (each vertex in the graph is mapped to one optical mode), this corresponds to a non-unitary process that maps orthogonal states to non-orthogonal states. Therefore, some non-unitary elements like post-selection or adaptive measurements should be properly added to the proposed framework, increasing the overall complexity of the implementation. A simpler hardware-tailored solution is to split the realization of the ECMs for stage 1 into independent circuits for each different percept. The random walk starting from the percept vertex to one of the four possible observable values vertices is realized with a single photon passing through a four-mode network of tunable beamsplitters arranged in a binary tree-like scheme (Fig. \ref{fig:qps} (b)). The input mode is kept fixed, and each of the output modes corresponds to a different value of the two observables. Each percept is associated with a different set of parameters, which are trained independently. 

In stage 2, a universal square mesh of tunable beamsplitters \cite{clements_optimal_2016} is added to the previous circuit to reproduce the “all-to-all" network among the four observable values and the last two possible answers. The set of parameters of this last part of the circuit is the set of trainable parameters, while the parameters of the first part are retrieved from the previous stage and swapped depending on the current percept. Again, the input mode is kept fixed, while at the output only the states with the photon in one of the first two modes ($\ket{1,0,0,0}$ and $\ket{0,1,0,0}$), corresponding to the “no"/“yes" answers, are postselected on.

Each circuit needed to realize the different stages of the quantum ECM can be implemented using Perceval \cite{heurtel_perceval_2023}, an open-source framework for linear optical quantum computing. In this work, three different backends are used: one to simulate the ideal performances (\emph{ideal}), one that accounts for the noise due to the imperfections of the single-photon source and due to photon loss (\emph{noisy}), and one to directly interface with the hardware (\emph{hardware}). The hardware used in this work is the Quandela 12 modes 6 photons photonic quantum processor Ascella \cite{maring_versatile_2024}. A detailed overview of \emph{ideal} and \emph{noise} can be found in Appendix B \cite{A_B}.

\section{Training procedure }
We now focus on how the predefined architecture can be trained to solve the task of interest. As mentioned before, the optimal set of parameters of the quantum ECM can be found using a variational approach. We train the agent by sequentially interacting with the environment at each stage of the learning scenario. After each interaction, or a predetermined set of interactions, the parameters are updated following the descending direction of the loss function gradient. Updating the parameters after each interaction is faster, but also less accurate; in stage 2 of the task, a better convergence is observed by updating the parameters after a set of 10 interactions. The training lasts for a given number of episodes, where one episode corresponds to one parameter update.

In stage 1, we train four independent sets of parameters (the tunable elements of the circuit underlined in yellow in Fig. \ref{fig:qps} (b)), one for every percept. An action sampling for each of them involves detecting the output mode of a single photon after it is input into the linear optical circuits of the ECM. Then, the agent is rewarded with $+(-) \,0.1$ if the chosen action does (not) correspond to one of the observable values of the current percept. The loss function to be optimized at this stage is the sum of three components:
\begin{equation} \label{loss_qps_1}
    \mathcal{L}_1(\vec{\theta}_1) = \mathcal{L}_{PS}(\vec{\theta}_1) + 10\mathcal{L}_{Shannon}(\vec{\theta}_1) + \mathcal{L}_{Phase}(\vec{\theta}_1).
\end{equation}
\begin{itemize}
\item 
The \emph{PS} component corresponds to \eqref{loss_fn}, where we take $C(x) = 0.5 - \text{ReLU}(0.5-\text{ReLU}(x))$ as the clamping function and the KL divergence \cite{kullback_information_1951} of the two distributions $\{p_{kl}^{t+1}(\vec{\theta}_1), 1-p_{kl}^{t+1}(\vec{\theta}_1)\}$ and $\{C(p_{kl}^{t}+r^t), 1-C(p_{kl}^{t}+r^t)\}$ as the distance function.
\item 
The \emph{Shannon} component relies on the Shannon entropy \cite{shannon_mathematical_1948} of the probability distribution 
$\{p_{color}, p_{shape} \}$, 
where $p_{o}$ is the probability of sampling an output mode corresponding to the observable $o$ (i.e. $p_{\emph{color}}$ is the probability of getting the photon in mode 0 or 1 at the end of the stage 1 ECM on Fig. \ref{fig:qps} (b)):
\begin{equation}
   \mathcal{L}_{Shannon} = \log{(2)} + \sum_{o \in \{\emph{color, shape}\}} p_o\log{(p_o)}.
\end{equation}
Here we are denoting $p_{o}^t(\vec{\theta_1})$ as $p_{o}$. This component works as a curiosity mechanism that motivates the agent to learn about both observables. A hyperparameter is employed to regulate the magnitude of this component with respect to the others, we adjust it to a value of $10$.
\item 
The \emph{Phase} component consists of the L$^1$ norm of the complex phases of the output state (at the end of the stage 1 ECM, Fig. \ref{fig:qps} (b)):
\begin{equation}
    \mathcal{L}_{Phase} = \sum _{m=0} ^{M-1} |\Phi_m|,
\end{equation}
where $\Phi_m$ is the phase on mode $m$. Minimizing this component allows to avoid unrestricted phases between different modes in the last layer of this stage. This feature becomes essential in stage 2, where the agent must exploit interference between optical modes to solve the task efficiently. 
\end{itemize}
The loss function of a set of agent-environment interactions is simply the sum of the respective $\mathcal{L}_1$ for each interaction. At this stage, the gradient is computed over batches of $10$ interactions using the simultaneous perturbation stochastic approximation (SPSA) algorithm \cite{spall_overview_1998}, and training is run for 400 episodes.

\begin{table*}[ht]
\small
\centering
\begin{minipage}{1\linewidth}
\centering
    \begin{tabular}{|c|P{1cm}|P{1cm}|P{1cm}|P{1cm}|P{1cm}|P{1cm}|P{1cm}|P{1cm}|P{1cm}|P{1cm}|P{1cm}|P{1cm}|}
                             \hline & \multicolumn{4}{c|}{\emph{ideal}}                      & \multicolumn{4}{c|}{\emph{noisy}}                & \multicolumn{4}{c|}{\emph{hardware} ($\pm 0.02$)}   \\\hline
                            \emph{percept} & $p_\text{red}$ & $p_\text{blue}$ & $p_\text{circle}$ & $p_\text{square}$ & $p_\text{red}$ & $p_\text{blue}$ & $p_\text{circle}$ & $p_\text{square}$ & $p_\text{red}$ & $p_\text{blue}$ & $p_\text{circle}$ & $p_\text{square}$ \\\hline
    0 & \cellcolor[HTML]{C0C0C0}0.50           & 0.00            & \cellcolor[HTML]{C0C0C0}0.49              & 0.01              & \cellcolor[HTML]{C0C0C0}0.50           & 0.00            & \cellcolor[HTML]{C0C0C0}0.49              & 0.01              & \cellcolor[HTML]{C0C0C0}0.47           & 0.01            & \cellcolor[HTML]{C0C0C0}0.5               & 0.02              \\
    1 & 0.02           & \cellcolor[HTML]{C0C0C0}0.49            & 0.01              & \cellcolor[HTML]{C0C0C0}0.48              & 0.01           & \cellcolor[HTML]{C0C0C0}0.49            & 0.01              &\cellcolor[HTML]{C0C0C0} 0.49              & 0.02           & \cellcolor[HTML]{C0C0C0}0.46            & 0.00              & \cellcolor[HTML]{C0C0C0}0.52              \\
    2  & \cellcolor[HTML]{C0C0C0}0.49           & 0.01            & 0.01              & \cellcolor[HTML]{C0C0C0}0.49              &\cellcolor[HTML]{C0C0C0} 0.49           & 0.01            & 0.01              &\cellcolor[HTML]{C0C0C0} 0.49              & \cellcolor[HTML]{C0C0C0}0.49           & 0.00            & 0.00              & \cellcolor[HTML]{C0C0C0}0.51              \\
    3 & 0.01           & \cellcolor[HTML]{C0C0C0}0.49            &\cellcolor[HTML]{C0C0C0} 0.49              & 0.01              & 0.02           &\cellcolor[HTML]{C0C0C0} 0.48            &\cellcolor[HTML]{C0C0C0} 0.49              & 0.01              & 0.01           &\cellcolor[HTML]{C0C0C0} 0.45            &\cellcolor[HTML]{C0C0C0} 0.54              & 0.00 \\\hline
    \end{tabular}\\ \vspace{0.15cm}
    (a)\vspace{0.15cm}
\end{minipage}

\begin{minipage}{1\linewidth}
\centering
 \begin{tabular}{|c|P{1.5cm}|P{1.5cm}|P{1.5cm}|P{1.5cm}|P{1.5cm}|P{1.5cm}|}
                    \hline & \multicolumn{2}{c|}{\emph{ideal}}                                  & \multicolumn{2}{c|}{\emph{noisy}}                            & \multicolumn{2}{c|}{\emph{hardware} $(\pm 0.02)$}                  \\ \hline 
    \textit{percept} & $p_\text{no}$                & $p_\text{yes}$               & $p_\text{no}$                & $p_\text{yes}$               & $p_\text{no}$                & $p_\text{yes}$               \\ \hline
    0                & \cellcolor[HTML]{C0C0C0}0.99 & 0.01                         & \cellcolor[HTML]{C0C0C0}0.97 & 0.03                         & \cellcolor[HTML]{C0C0C0}0.95 & 0.05                         \\
    1                & 0.01                         & \cellcolor[HTML]{C0C0C0}0.99 & 0.00                         & \cellcolor[HTML]{C0C0C0}1.00 & 0.00                         & \cellcolor[HTML]{C0C0C0}1.00 \\
    2                & \cellcolor[HTML]{C0C0C0}0.99 & 0.01                         & \cellcolor[HTML]{C0C0C0}0.99 & 0.01                         & \cellcolor[HTML]{C0C0C0}0.99 & 0.01                         \\
    3                & \cellcolor[HTML]{C0C0C0}1.00 & 0.00                         & \cellcolor[HTML]{C0C0C0}0.99 & 0.01                         & \cellcolor[HTML]{C0C0C0}0.99 & 0.01          \\\hline              
    \end{tabular}
    \\ \vspace{0.15cm}
    (b)
\end{minipage}

\caption{\textbf{Probability distributions of the trained quantum EMC for stage 1 (a) and 2 (b).} The output modes probabilities are mapped to the probabilities of the values of the observable. The correct values for each percept are highlighted in grey.}
\label{tab:mid_lay}
\end{table*}

In stage 2, we train only the last part of the circuit that was added on top of the stage 1 ECM. A different percept is sampled uniformly at each interaction, and the loss function considered corresponds to that in \eqref{loss_qps_1} with only the \emph{PS} component. If the chosen action is correct, the agent receives a positive reward of $+1$, otherwise a negative reward of $-0.5$. The asymmetric choice of rewards is justified by the asymmetric nature of this last stage, where sampling a percept for which the correct answer is “yes" is less likely than sampling a percept for which the correct answer is “no". The gradient is computed after each interaction using the Finite Difference Stochastic Approximation (FDSA) method, and a total of 1000 training steps are considered. FDSA is a slower but more precise method for computing the approximate gradient compared to SPSA. We observed better training performance for the agent with FDSA at this stage. This may be due to the fact that we are optimizing over a larger set of parameters compared to stage 1, where SPSA was sufficient.

\section{Results }
When running the experiments to train the agent, the following pipeline is used at each stage: first, the experiments are run on the ideal simulator, then the training is repeated on the noisy simulator, and finally on the real hardware. The tuning of the hyperparameters of the optimizers is done in the first two stages of the pipeline. The ideal simulator is used for a wide search in the hyperparameter space and then a narrower set of possibilities is explored on the noisy simulator. Once the optimal set of hyperparameters is chosen, the training is performed again with the ideal simulator to allow for better comparison between backends. This procedure is chosen to ensure successful training when running on the hardware.

\begin{figure}[b!]
\includegraphics[width=0.4\textwidth]{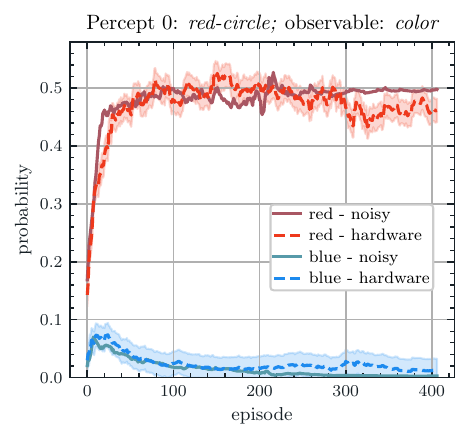}
\caption{
\label{fig:p_0} \textbf{Evolution of the \emph{color} observable probability distribution during stage 1 training.} The graph shows the probability distribution of the observable \emph{color} during stage 1 of training for the quantum ECM corresponding to percept 0 \emph{red-circle}. The horizontal axis displays the number of episodes within the training.}
\end{figure}

The final probability distribution after stage 1 training is shown in Tab. \ref{tab:mid_lay} (a), already encoded in the values of the observables. We report the results for the three backends; those relative to \emph{hardware} experiments have an associated error bar, which is discussed in Appendix C \cite{A_C}. The evolution of the probabilities for the observable \emph{color} of percept 0 is shown in Fig. \ref{fig:p_0}. As expected, at the end of training, given a percept, the agent can select the correct color in half of the cases and the correct shape in the other half of the cases. The final \emph{hardware} output probabilities are slightly worse compared to the other backends, however, taking into account the uncertainty associated, they are still compatible with the expected ones.

Table \ref{tab:mid_lay} (b) shows the final probability distribution after stage 2 of the learning task, already taking into account the postselection on the first two modes of the circuit and the encoding of the optical modes into “no"/“yes" possible answers. As mentioned above, the goal here is to correctly answer the query \emph{is the percept both blue and a square?}. Once trained, the agent achieves the correct behavior by giving a positive answer only for the \emph{blue-square} percept and a negative answer in the other cases. This happens for all backends, although results closer to the target distribution are achieved in the ideal simulator case, as expected.

\begin{figure}[]
\includegraphics[width=0.4\textwidth]{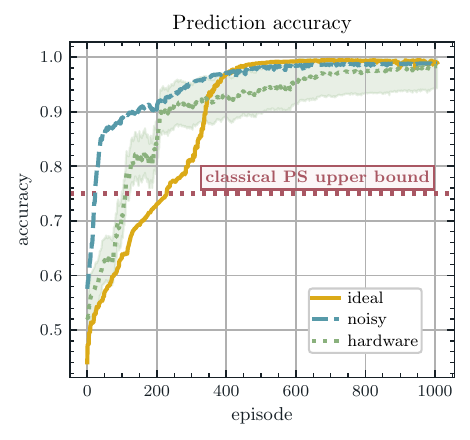}
\caption{\label{fig:acc} \textbf{Evolution of accuracy score during stage 2 training.} The graph shows the accuracy in solving the task during the training of stage 2 of the quantum ECM. The horizontal dotted line shows the classical ECM accuracy upper bound. The accuracy is displayed as a percentage along the vertical axis, while the number of training episodes is represented on the horizontal axis.}
\end{figure}

Next, we aim to apply the knowledge acquired in stage 1 to accurately respond to the query \emph{is the percept both blue and a square?} At each step of the training for this stage, we compute the accuracy of the agent solving the task as follows:
\begin{equation} \label{eq:acc}
    \text{accuracy} = \frac{1}{4} \sum _{i=0} ^3 \left( p_\text{no} ^i \cdot \Lambda ^i _{\text{no}} + p_\text{yes} ^i \cdot \Lambda ^i _{\text{yes}} \right),
\end{equation} 
where $p_\text{yes/no} ^i$ is the probability with which the agent answers “yes"/“no" given percept $i$.
The value of $ \Lambda ^i _{\text{no}} $ is 1 if $ i $ is in the set \{0, 2, 3\}, and 0 otherwise. Similarly, the value of $ \Lambda ^i _{\text{yes}} $ is 1 if $ i $ is in the set \{1\}, and 0 otherwise. The accuracy of the different backends over the episodes is shown in Fig. \ref{fig:acc}. 
This plot can be understood as a classical RL score function for the agent, in our case approaching $100\%$ accuracy means successfully solving the task. The dotted line represents the maximum accuracy that a classical PS agent can achieve with the corresponding classical ECM. The different performances in terms of the evolution of the accuracy of the three backends are mainly related to the transfer learning nature of the task being tackled. The fact that the \emph{ideal} accuracy converges to $100\%$ slower than the other two backends is due to the different set of parameters found at the previous stage of the task. Even though the same optimization landscape is explored, different starting points lead to different trajectories toward the minima of the loss function.

\section{Discussion and conclusion }
In this work, we presented the first physical realization of a quantum PS agent on a photonic quantum processor. The core of our contribution was to tune both the original learning task and the training procedure to the hardware at our disposal. We showed that the quantum agent can achieve full accuracy in solving the task on an actual noisy device, thus complementing the results of \cite{Flamini_2024} obtained on an ideal simulator. The superiority of the quantum agent with respect to its classical counterpart holds for this specific transfer learning scenario when the main structure of both agents, i.e. the graph underlying the ECM, is the same. Even if the scenario in question is simple, it serves well both the purpose of providing a test bed task for the hardware implementation of the quantum PS agent, and the purpose of showing in which kind of scenario the quantum properties of the agent begin to play a relevant role. Given this starting point, one could try to create more complex and interesting tasks where the hardware requirements are feasible and the agent's quantum properties provide some potential.

One avenue for future research is to extend our work to the multi-photon setting. Instead of taking decisions through single-photon quantum walks, multiple photons quantum walks can be evaluated to pick an action. In the classical setting, this would correspond to simultaneous multiple deliberations within the ECM in the decision process. The hardware requirements in this case are not a significant concern, as the QPU used in this work is already capable of handling up to six photons at the input stage. However, the theory supporting this approach is still under development, with some progress made in extending the classical PS algorithm to a multi-excitation setting \cite{lemaitre2024multi}. Another possible direction would be to implement a variant of the PS algorithm, the reflecting PS algorithm (r-PS). The r-PS, implies that the deliberation process is repeated many times, and then the choice of the action is dictated by the probability distribution over the clips of the ECM. For this variant, it has been shown that there is the presence of a quadratic speed-up for the quantum agent with respect to the classical counterpart in terms of the learning time \cite{paparo_quantum_2014}. Therefore, a promising direction for the extension of this work is to investigate the implementation of the reflective variant of PS on a photonic platform. As a final consideration, when scaling to more complex tasks, both in the version of the algorithm analyzed here and in possible extensions, the main experimental bottleneck seems to be related to the number of optical modes of the device, since a number of modes equal to the dimensionality of either the percepts space or the space of observable values is required, depending on the nature of the task.

Overall, by systematically evaluating the quantum PS algorithm in a controlled setting, our study contributes to a broader understanding of the applicability of VQAs to machine learning tasks on photonic quantum devices.\\

The trained parameters for both stages of the learning scenario are available at \cite{quantum_PS}, along with a short tutorial on how to upload them to the corresponding linear optical circuits using Perceval.

\section*{Acknowledgments }
We thank Marius Krumm for fruitful discussions throughout the project. We thank Alexia Salavrakos and Pierre-Emmanuel Emeriau for their careful revision of the paper. A.R. and G.F. acknowledge funding from AQGeTAD; PAQ programme from Ile-de-France Region. A.R. acknowledges funding from the European Commission as part of the EIC accelerator program under the grant agreement 190188855 for SEPOQC project. G.F. acknowledges support from a ”la Caixa” Foundation (ID 100010434) fellowship. The fellowship code is LCF/BQ/DI23/11990070.

\section*{Appendix A: Upper bound on the accuracy of the classical PS agent}

In this section, we show how to derive the accuracy upper bound of the classical PS agent in solving the learning task presented in this work. At the end of the first stage, it is reasonable to assume that the agent is able to achieve a perfect accuracy in characterizing each percept. Consequently, the initial condition for the second stage is an ECM, where, given the excitation of a single percept clip, the decision path will proceed to the clip of the correct color in half of the cases and the clip of the correct shape in the other half of the cases. Thereafter, within stage 2 the environment begins to sample the different percepts in a uniform manner, and the agent is faced with the task of correctly answering the query \emph{is the percept both blue and a square?} For the agent, picking a decision means going through the action clips, thus either the “yes" clip or the “no" clip. At this point, we claim that the agent cannot reach a total accuracy of over $75\%$. Here, we denote the \emph{blue-square} percept as the target percept and we say that a percept has overlapping observable values with another percept if two percepts have the color and/or the shape in common. We proceed to analyze the accuracy contributions for each percept:
\begin{itemize}
\item \emph{Red-circle (no overlapping observable values with the target percept)} - In this case, the decision path proceeds through either the clip red or the clip circle. For both clips, the only successive connection to be positively rewarded is the one with the no clip, which is the correct answer. Since the current answer remains “no" also when these intermediate clips are reached from different percepts, the agent can learn to answer with $100\%$ accuracy for this percept.
\item \emph{Blue-circle, red-square (one overlapping observable value with the target percept)} - Two possibilities exist in this context. When the decision path traverses an observable value that is not overlapping with the target percept (half of the cases), the agent can learn to correctly answer “no" with $100\%$ accuracy, as previously described. On the other hand, when the decision path traverses an observable value that overlaps with the target percept (other half of the cases), the situation is different. For these percepts, only the connection with the “no" clip is positively rewarded. However, whenever the target percept is sampled, this same connection is negatively rewarded since the correct answer is “yes" for this case. At the end of training, the optimal strategy for the agent involves randomly picking one of two action clips. Therefore, this results in a final accuracy of $75\%$ for these two percepts.
\item \emph{Blue-square} - For the target percept, we revert to the scenario that we explored for the overlapping observable values of the previous percepts. The connection to answer correctly is going to be positively or negatively rewarded, depending on the given percept. Consequently, a maximum accuracy of $50\%$ can be achieved at the end of the training.
\end{itemize}
By summing the contributions of all percepts, it is possible to compute the maximum accuracy that a classical agent can achieve. This turns out to be $\frac{1}{4}\cdot100\%+\frac{1}{2}\cdot75\%+\frac{1}{4}\cdot50\% = 75\%$.

\section*{Appendix B: Perceval's backends}
Perceval is an open-source framework to simulate ideal and noisy linear optical circuits that also allows to connect and run experiments on real quantum photonic devices connected to the cloud, such as Ascella. Within this framework, different backends to perform both strong and weak simulations of photonic circuits, are available. Strong simulations compute the exact complete output state of the circuit while weak simulations produce samples from the output probability distribution. To allow for a better understanding of the content, within the main text, we omitted the technical names of the different backends and referred to them as \emph{ideal, noisy} and \emph{hardware}. Here we make use of the technical names, the \emph{ideal} backend corresponds to \texttt{SLOS}, the \emph{noisy} one to \texttt{sim:ascella}, and the one interfacing with the hardware \texttt{qpu:ascella}. A short description of the functioning of \texttt{SLOS} and \texttt{sim:ascella} is provided below.

\subsubsection{\texttt{SLOS}}
The \emph{strong linear optical simulation} (\texttt{SLOS}) algorithm, introduced in \cite{heurtel_strong_2023}, consists of a strong ideal simulation backend that unfolds and optimizes the full computation of the output probability distribution of the circuit. From \cite{aaronson_computational_2010}, we know that the probability of getting the $n$-photon output state $\ket{O} = \ket{o_1, \dots, o_m}$ given a circuit that implements the $m \times m$ unitary $\mathcal{U}$ and the $n$-photon input state $\ket{I} = \ket{i_1, \dots, i_m}$ is related to the permanent of a matrix $U_{O,I}$. $U_{O,I}$ is a $n \times n$ matrix that is derived in two steps:
\begin{enumerate}
    \item First, we form a $m \times n$ matrix $U_I$ taking $i_j$ copies of the $j$-th column of $\mathcal{U}$ for each $j \in \{1, \dots, m\}$.
    \item Then, we form $U_{O,I}$ by taking $o_k$ copies of the $k$-th row of $U_I$ for each $k \in \{1, \dots, m\}$.
\end{enumerate}
The fastest algorithms \cite{ryser_combinatorial_1963,glynn_permanent_2010} to compute the permanent of a $n \times n$ matrix are in $\mathcal{O}(n2^n)$. \texttt{SLOS} exploits the fact that $U_{O,I}$ is a matrix with repeated rows or columns to reduce the complexity of computing its permanentes to $\mathcal{O}(n\binom{n+m-1}{m-1})$. This backend is used to simulate the expected ideal behavior of linear optical circuits by computing the probabilities of all the possible output states. \\
\subsubsection{\texttt{sim:ascella}}
The \texttt{sim:ascella} backend is a strong simulator that is used to emulate the precise behavior of Ascella. It is hosted on the cloud and takes into account both the compilation process and some imperfections of the physical device. The compilation process is performed by the same algorithm that is used to compile the circuits for the real device. Instead of simulating the circuit received as input, this backend simulates the circuit that is actually implemented on the device. As of the device imperfections, it takes into account:

\begin{itemize}
    \item $g^{(2)}(0)$ factor, that corresponds, at the first order of approximation, to the probability of emitting two photons when one was expected. Then, $1- g^{(2)}(0)$ matches the single photon source purity. Typical values of $g^{(2)}(0)$  for Ascella are of the order of $10^{-3}$.
    \item HOM interference visibility $V_{HOM}$, that, together with $g^{(2)}(0)$, is used to estimate the single photon source 2-photon indistinguishability \cite{ollivier_hong-ou-mandel_2021}. $V_{HOM}$ ranges from $0$ to $1$, the higher the value, the higher the indistinguishability. Standard values of $V_{HOM}$ for Ascella are $ > 0.9$.
    \item The total transmission of the setup, that is characterized by the efficiency of each module of the setup, starting from when the photon is generated until its detection. Typical values for Ascella are around the $8 \%$.
\end{itemize}

The technique that allows to take into account these imperfections is based on a phenomenological model of the source, well presented in \cite{pont_high-fidelity_2022}. This model provides all the imperfect states that could be emitted by the single photon source. Then, the statistical mixture of these states is considered as input. Moreover, assuming that the transmission factor affects equally each mode, its effect can be commuted with all the other optical elements and included in the input state \cite{garcia-patron_simulating_2019}. Finally, an ideal simulation with the \texttt{SLOS} backend is performed and all the compatible output states are collected and normalized.

\section*{Appendix C: Average variance on Ascella}

When training the quantum PS agent on the actual device, we have to select the number of shots with which we need to sample the circuit to get a good estimation of the output distribution. In fact, while with the ideal backend \texttt{SLOS} we can access the full output distribution of the circuit, with Ascella, we can only get an estimation of that by sampling the circuit multiple times. Hence, the number of shots, $n.o.s.$, that we use to sample the circuit is a parameter that we have to properly tune.\\

\begin{figure}[!ht]
    \centering
    \includegraphics[width=0.48\textwidth]{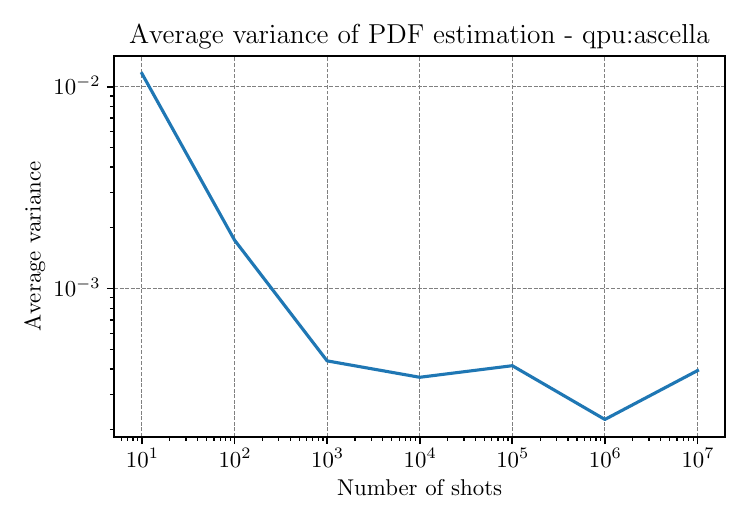}

    \caption{Average variance of the output distribution for different $n.o.s.$ on \emph{qpu:ascella}.}
    \label{fig:avg_var}
\end{figure}

To evaluate the effect of the number of shots on the estimation of the output distribution, we consider the tree-like circuit used in stage 1 of the learning task and we compute the average variance of the output distribution computed for different $n.o.s.$ as follows:
\begin{widetext}
    \begin{equation}
    \langle \sigma ^2 _{\emph{qpu:ascella}} \rangle = \frac{1}{N} \sum _{i=1} ^N \frac{1}{4}\sum _{j=1} ^4 \left(  p(a_j, n.o.s.|a_k)_{\emph{qpu:ascella}}- p(a_j, n.o.s.|a_k)_{\emph{SLOS}}\right) ^2,
\end{equation}
\end{widetext}
where $N=10$ and $p(a_j, n.o.s.|a_k)_{\emph{backend}}$ is equal to the estimated (or exact in the case of \texttt{SLOS}) probability of getting the output mode $c_j$ after $n.o.s.$ runs of the circuit with input mode $a_k$, using the given \emph{backend}. In Figure \ref{fig:avg_var} we show the average variance of the output distribution versus different $n.o.s.$ for \texttt{qpu:ascella}. 
As expected the variance tends to decrease with the number of shots. However, we see that for $n.o.s. \geq 10^4$ the improvement trend slows down. Therefore, for this region of the plot, it seems that a saturation regime is reached in terms of average variance. Even if the difference is small, from the extracted data the optimal number of shots seems to be $n.o.s. = 10^6$. Despite that, we decide to use $n.o.s. = 10^5$ for the training of the agent. This choice is motivated by the fact that the training of the agent is a time-expensive task and we believe that the increase in performance with $n.o.s. = 10^6$ or higher with respect to $10^5$ is not worth the additional time required. \\

At this point, we also avoid selecting $10^4$ shots even if this value has a similar average variance as $10^5$ and would mean faster evaluations of the circuits. To explain this choice, another detail has to be taken into account. Even if we can ask for a little number of shots from a circuit on \texttt{qpu:ascella} we have to consider the physical limitations of the device. In fact, the single photon source emits photons at a rate of $\SI{80}{\MHz}$, the demultiplexer then delays the produced photons to be able to give input states with $6$ coincident photons. When asking for single photon input states, 5 out of the 6 input lines are blocked but still the input rate that you obtain is of $\SI{80}{\MHz}/6 = \SI{13.3}{\MHz}$. On top of that, output data are collected every second and we have to consider the losses throughout the device. This sums up into getting at least $\sim 10^5$ samplings from \texttt{qpu:ascella} every single photon sampling job that is launched, even if we require just a single shot. This justifies why we preferred to use $n.o.s. = 10^5$ instead of $n.o.s. = 10^4$. 

With this analysis, not only we properly tune the number of shots for the training of the agent on Ascella, but we also get an estimate of the standard deviation of the output probabilities that we obtain from the device. Hence, the \texttt{qpu:ascella} results that we present are reported with error bars equal to $2\sqrt{\langle \sigma ^2 _{\emph{qpu:ascella}} (10^5) \rangle} \simeq 2 \cdot 0.02$.

\bibliography{main}

\end{document}